\documentclass[CJK,CCT,openany,twoside]{cctbook}  
\usepackage{graphicx}
\usepackage{amssymb,vatola} 
\usepackage{multicol}  
\usepackage{fancyhdr}
\usepackage{amssymb}
\usepackage{vatola}
\pdfoutput=1
\input cyracc.def
\DeclareGraphicsExtensions{.epsi,.eps,.ps,.eps.gz,.ps.gz,.gif,.epsi.gz}
\usepackage{color}
\usepackage{xspace}
\usepackage{bm}                       

\newcommand{\babar}{BaBar\xspace}

\def\Dz{\ensuremath{D^0}\xspace}
\def\Dzb{\ensuremath{\overline{D}^0}\xspace}

\def\invfb{\ensuremath{\mathrm{fb}^{-1}}\xspace}
\def\invab{\ensuremath{\mathrm{ab}^{-1}}\xspace}

\newfont{\yihao}{cmb10 at 18pt}

\usepackage{calc}                     

\newenvironment{OOExercises}[1][10..]{
\begin{list}
{\hfill\arabic{enumi}.}{
\settowidth{\labelwidth}{\bfseries #1}%
\setlength{\labelsep}{8pt}%
\setlength{\topsep}{10bp}%
\setlength{\partopsep}{5pt}%
\setlength{\parskip}{0pt}%
\setlength{\itemsep}{0bp}%
\setlength{\leftmargin}{\labelwidth+\labelsep}%
\usecounter{enumi}}}{\end{list}
}

\DeclareSymbolFont{lettersA}{U}{txmia}{m}{it}
\DeclareMathSymbol{\piup}{\mathord}{lettersA}{25}
\DeclareMathSymbol{\muup}{\mathord}{lettersA}{22}      
\DeclareMathSymbol{\gammaup}{\mathord}{lettersA}{13}      

\renewcommand{\baselinestretch}{1.06} 
 \catcode`@=11

{%
\fancyhead{} 

}

\definecolor{orangec}{cmyk}{.24,.91,.96,.18}

\definecolor{orangecc}{cmyk}{.24,.94,.96,.18}
\definecolor{oorangec}{cmyk}{.8,.2,.5,.4}
\definecolor{ooorangec}{cmyk}{1,.9,0.08,.04}      

\renewcommand\footnoterule{\vspace*{3pt}
\hrule height 0pt \vspace*{3pt}}

\newfont{\xbt}{cmb10 at 12pt}

\usepackage{graphicx}                 

\graphicspath{{figs/}}      

\usepackage{xspace}

\def\C{\ensuremath{C}\xspace}
\def\P{\ensuremath{P}\xspace}
\def\T{\ensuremath{T}\xspace}
\def\CP{\ensuremath{CP}\xspace}
\def\CPT{\ensuremath{CPT}\xspace}

\def\epem{\ensuremath{e^+e^-}\xspace}

\def\mev{\ensuremath{\mathrm{MeV}}\xspace}

\begin{document}
\thispagestyle{empty}



\begin{center}

{\usefont{T1}{fradmcn}{m}{n}\yihao Testing discrete symmetries at a super \boldmath {\Large $\tau$}-charm factory} 

{\bf\small Adrian John Bevan}\vspace*{4mm}

{\footnotesize\it
 Particle Physics Research Centre, Queen Mary University of London, Mile End Road, London, E1 4NS, UK\\[1.5mm]
    E-mail: a.j.bevan@qmul.ac.uk\\[1mm]

 }\vspace{5mm}

\baselineskip 10pt
\renewcommand{\baselinestretch}{0.8}
\parbox[c]{152mm}
{\noindent{{\color{oorangec}
Tests of discrete symmetry violation have played an important role in understand the structure of weak interactions in the Standard Model of particle
physics.  Historically these measurements have been extensively performed at experiments with large samples of $K$ and $B$ mesons.  A high luminosity
$\tau$-charm facility presents physicists with the opportunity to comprehensively explore discrete symmetry violation and test the 
Standard Model using $\tau$ leptons, charm mesons and charmed baryons.  This paper discusses several possible measurements for a future
$\tau$-charm factory.
}\vspace{2mm}

{\color{ooorangec}\bf Keywords}~~Discrete symmetries, \P, \C, \T \CP, \CPT violation\vspace{2mm}

{\color{ooorangec}\bf PACS numbers}~~14.40.Gx, 12.39.Mk, 13.20.He }}
\end{center}
\normalsize

\baselineskip 12pt \renewcommand{\baselinestretch}{1.06}
\parindent=10.8pt  \parskip=0mm \rm\vspace{2mm}

\begin{multicols}{2}
\setlength{\parindent}{1em}    

\centerline{Contents}\vspace{2.5mm}

\noindent 1\quad  Introduction\hfill 1\vspace{0.0mm}

\noindent 2\quad Triple product asymmetry measurements\hfill 1\vspace{0.0mm}\\
\noindent 2.1\quad $K_{L, S}\to\pi^+\pi^-e^+e^-$\hfill 2\vspace{0.0mm}\\
\noindent 2.2\quad Testing charm mesons and baryons\hfill 3\vspace{0.0mm}\\
 \noindent 2.2.1\quad Charm mesons decays\hfill 3\vspace{0.0mm}\\
 \noindent 2.2.2\quad Charm baryon decays\hfill 4\vspace{0.0mm}\\
\noindent 2.3\quad $\tau$ decays\hfill 4\vspace{0.0mm}\\
\noindent 2.4\quad A simple model\hfill 5\vspace{0.0mm}

\noindent 3\quad Tests using entangled states\hfill 5\vspace{0.0mm}\\
\noindent 3.1\quad Using flavour filters\hfill 6\vspace{0.0mm}\\
\noindent 3.2\quad Using \CP filters\hfill 7\vspace{0.0mm}\\
\noindent 3.3\quad Using both flavour and \CP filters\hfill 7\vspace{0.0mm}\\
\noindent 3.4\quad Tests of quantum mechanics\hfill 7\vspace{0.0mm}

\noindent 4\quad Summary\hfill 8\vspace{0.0mm}

\hspace{1.5mm}References and notes\hfill 8\vspace{0.0mm}

%
%
\vspace*{7mm} {\color{ooorangec}\hrule}\vspace{2mm} \noindent
{\color{ooorangec}\large
\usefont{T1}{fradmcn}{m}{n}\xbt 1\quad Introduction}\vspace{3.5mm}

\noindent 
The Standard Model of particle physics (SM) describes weak, strong and electromagnetic interactions.  The weak interaction is
known to violate the discrete symmetries \C, \P, \T and the combination \CP.  The combination \CPT is observed to be conserved.  
Strong and electromagnetic interactions conserve these symmetries. Parity violation was discovered
in 1957 [1] and \CP violation a few years later in 1964 [2].  Following these discoveries there was an interest in trying to
validate \T independently of \CPT. Whilst it was recognised that \CPT conservation was desirable given the prior evidence 
available, it was noted that testing \T independently of \CPT was important [3].  It is possible to test the full set of 
discrete symmetries using triple product asymmetries and using entangled pairs of neutral mesons.  This paper discusses
the potential for a Super $\tau$-charm facility in terms of testing discrete symmetries using $\tau$ leptons, charm mesons
and charm baryons produced near threshold.  A number of routes toward \CP violation measurements in charm decays 
are under study in the literature and have been discussed at length elsewhere, for example Refs [4-6]; 
here we review additional possibilities 
to probe discrete symmetries that complement the traditional routes. 

The remainder of this paper discusses the use of triple product asymmetries with four body decays to test \C, \P and \CP (Section 2),
followed by the use of entangled pairs of $D$ mesons produced in the decay of $\psi(3770)$ mesons to test \CP, \T and \CPT (Section 3).
Finally Section 4 presents a summary of this paper.  The data sample assumed for a Super $\tau$-charm facility is 1\invab, which 
corresponds to $10^9$ $\psi(3770)$ ($10^8$ $\psi(4040)$) mesons for $D^{0, \pm}$ ($D_s^\pm$) pair production.  Facilities
capable of producing these kinds of sample sizes are under investigation, for example the proposed High Intensity Electron Positron Accelerator (HIEPA)
in China.

\renewcommand\footnoterule{\vspace*{3pt}
\noindent
\def\temptablewidth{\textwidth}{\rule{0.3\temptablewidth}{0.5pt}}\vspace*{3pt}}

%
%
\vspace*{7mm} {\color{ooorangec}\hrule}\vspace{2mm} \noindent
{\color{ooorangec}\large
\usefont{T1}{fradmcn}{m}{n}\xbt 2\quad Triple product asymmetry measurements }\vspace{3.5mm}

\noindent 
If one considers the decay of some particle $M$ to a four body final state $abcd$ and the \CP conjugate process
$\overline M \to \overline{abcd}$, then it is possible to use the decay planes defined by the four vectors (or spins)
of pairs of final state particles to construct a scalar triple product that allows us to probe the symmetry violating
nature of the decay, for example see Refs [7,8].  The scalar triple product can be written as
$\psi = \vec{p}_c \cdot (\vec{p}_a \times \vec{p}_b)$, where the $\vec{p}_i$, $i=a,b,c$ 
are particle momentum vectors computed in the rest frame of $M$.  We can study data in terms
of the sign of $\psi$, or as a function of the angle between the decay planes formed by $ab$ and $cd$ in the reference 
frame of the decaying particle; $\phi$.  The angle $\phi$ is used when the underlying amplitudes in the 
decay are known sufficiently well to allow experimenters to understand if the interesting asymmetries are functions of
$\sin\phi$ or $\sin 2\phi$.  A number of measurements have been made in terms of the sign of the triple
product.  Following this generic approach we define $\Gamma_\pm$ to be the rate at which $M$
decays to a state with $\psi > 0$ ($+$) or $<0$ ($-$).  The corresponding rates for antiparticles
are given by $\overline{\Gamma}_\pm$.

Twelve asymmetries can be constructed by considering $\Gamma_\pm$ and $\overline \Gamma_\pm$ [8].  The first
six are derived by considering the \P, \C and \CP operators acting on the four $\Gamma$s.  These yield:
\begin{eqnarray}
A_{\P} = \frac{\Gamma_{+} - \Gamma_{-}} {\Gamma_{+} + \Gamma_{-}}, &\,\,\,\,\,&
\overline{A}_{\P} = \frac{\overline\Gamma_{+} - \overline\Gamma_{-}} {\overline\Gamma_{+} + \overline\Gamma_{-}},\label{eq:tp:parity}\\
A_{\C} = \frac{\overline\Gamma_{-} - \Gamma_{-}} {\overline\Gamma_{-} + \Gamma_{-}}, &\,\,\,\,\,&
\overline A_{\C} = \frac{\overline\Gamma_{+} - \Gamma_{+}} {\overline\Gamma_{+} + \Gamma_{+}},\label{eq:tp:c}\\
A_{\CP} = \frac{\overline{\Gamma}_{+} - \Gamma_{-}} {\overline{\Gamma}_{+} + \Gamma_{-}}, &\,\,\,\,\,&
\overline{A}_{\CP} = \frac{\overline\Gamma_{-} - \Gamma_{+}} {\overline\Gamma_{-} + \Gamma_{+}}.\label{eq:tp:cp}
\end{eqnarray}
Here the subscript indicates the symmetry being tested.  One can construct an additional six asymmetries considering
the remaining permutations, where the superscript denotes the original symmetry considered and the subscript denotes
the subsequent permutation:
\begin{eqnarray}
a_{\C}^{\P}&=& \frac{1}{2}\left( A_{\P} - \overline{A}_{\P}\right), \nonumber\\
a_{\CP}^{\P} &=& \frac{1}{2}\left( A_{\P} + \overline{A}_{\P}\right),\nonumber\\
a_{\P}^{\C} &=& \frac{1}{2}(A_{\C} - \overline{A}_{\C}),\nonumber\\
a_{\CP}^{\C} &=& \frac{1}{2}(A_{\C} + \overline{A}_{\C}),\nonumber\\
a^{\CP}_{\P} &=& \frac{1}{2}(A_{\CP} - \overline A_{\CP}),\nonumber\\
a^{\CP}_{\C} &=& \frac{1}{2}(A_{\CP} + \overline A_{\CP}).\label{eq:tp:derived}
\end{eqnarray}
The symmetry being tested by these last six asymmetries can be determined by multiplying the sub- and super-scripts together.
There are three types of decay that we can consider measuring, the most general case has been considered so far, however
we can consider the simplification when $abcd=\overline{abcd}$.  In this limit the twelve asymmetries remain non-trivial.
In the case that we further simplify to also require that $M=\overline M$ we obtain only a single unique and non-trivial asymmetry
given by
\begin{eqnarray}
A_{\P, \CP} = \frac{\langle\Gamma\rangle_{+} - \langle\Gamma\rangle_{-} }{ \langle\Gamma\rangle_{+} + \langle\Gamma\rangle_{-}},\label{eq:tp:averagedratescp}
\end{eqnarray}
where the average rates are indicated to highlight that $M$ is indistinguishable from $\overline M$. 
Before discussing charm mesons it is useful to proceed via an interlude (Section 2.1)
that reviews triple product asymmetry measurements in neutral kaon decays.  Following this we discuss
applications to charm mesons and baryons (Section 2.2) and $\tau$ leptons (Section 2.3).
We discuss a model-based interpretation of these asymmetries in Section 2.4.  

\vspace*{7mm} {\color{ooorangec}\hrule}\vspace{2mm} \noindent
{\color{ooorangec}\large
\usefont{T1}{fradmcn}{m}{n}\xbt 2.1\quad  $K_{L, S}\to\pi^+\pi^-e^+e^-$}\vspace{3.5mm}

The decay $K_L\to\pi^+\pi^-e^+e^-$ has been studied both theoretically and experimentally.  Reflection on these
results provides useful insight into how to address measurements of triple product asymmetries in the charm sector.
It was noted by Heiliger and Sehgal that this mode proceeds via four amplitudes; $K_L\to\pi^+\pi^-\gamma$ photon conversion;
bremsstrahlung from the \CP violating decay $K_L\to\pi^+\pi^-$; a \CP conserving magnetic dipole component;
and finally a short distance component related to $s\overline{d}\to e^+e^-$.  The radiative $K_L\to
\pi^+\pi^-$ decay is \CP violating and it is the interference between this amplitude and the remaining \CP conserving
ones that gives rise to a non-zero \CP asymmetry.  Heiliger and Sehgal predicted that the level of \CP violation
manifest in this decay is of the order of 14\% [9].  Shortly after this prediction was made the KTeV experiment
at the Fermi National Accelerator Laboratory measured this triple product asymmetry and confirmed the existence of 
a large effect [10].  Subsequently the NA48 experiment measured the triple product asymmetry of both the 
$K_L$ and $K_S$ meson decaying into $\pi^+\pi^-e^+e^-$ [11].  These results were found to be 
consistent with KTeV for the $K_L$ mode,
and consistent with \CP conservation for the $K_S$ decay as expected (given that $K_S\to \pi^+\pi^-$ is \CP conserving).  
This highlights an important issue with 
regard to \CP asymmetries; first one needs to identify a \CP violating amplitude, and only then the interference of 
that amplitude with other contributions may manifest effects that will be non-zero.  This is a well known 
statement of fact and is far from being profound. This factor should be taken 
into consideration during the following discussion with regard to possible measurements that one can make, and 
measurements that have been made. Thus far $D$ decays to four body final states studied 
require amplitude analyses for a complete interpretation of results.  These 
are complicated and have not yet been attempted; a priori it is not clear from 
inclusive measurements if non-zero triple product asymmetries are driven (in part) by a
non-zero weak phase difference between pairs of amplitudes or not.  
One has to understand the dominant amplitude contributions
to the decay model and from that model one can evaluate what the expected outcome might be.  At the time of 
writing model dependent analyses have not been performed, however a simple example is discussed below in Section 2.4.
It is hoped that measurement (and theoretical considerations) will evolve to permit model dependent studies of
these decays over the coming decade.

\vspace*{7mm} {\color{ooorangec}\hrule}\vspace{2mm} \noindent
{\color{ooorangec}\large
\usefont{T1}{fradmcn}{m}{n}\xbt 2.2\quad Testing charm mesons and baryons }\vspace{3.5mm}

In section 2.2.1 we start by considering tests using charm mesons with a brief summary of the current state of the art in terms of measurements, 
then move on to discuss possible measurements for the future.  In doing so we link back to reflect on the 
work done in kaon decays to draw analogies and highlight several modes that have been ignored thus far. 
Having discussed
measurements with mesons we move to consider baryon (Section 2.2.2) and $\tau$ lepton (Section 2.3) decays.

\vspace*{7mm} {\color{ooorangec}\hrule}\vspace{2mm} \noindent
{\color{ooorangec}\large
\usefont{T1}{fradmcn}{m}{n}\xbt 2.2.1\quad Charm meson decays }\vspace{3.5mm}

The most studied triple product asymmetries for four body $D$ decays are for 
the channel $D^0 \to K^+K^-\pi^+\pi^-$.  This has been studied by FOCUS, \babar and
LHCb and provides an interesting window of opportunity given a relatively large
branching ratio; $(2.43 \pm 0.12)\times 10^{-3}$.  
Experimentally the symmetry in the final state results in cancellation
of a number of systematic uncertainties. The FOCUS measurements were insufficient
to establish any non-zero triple product asymmetry, but laid the foundations for subsequent
work by \babar; this $B$ factory initially repeated the FOCUS measurement but with a larger 
data sample.  \babar found
non-zero values for $A_P$ and $\overline A_P$, but the \CP asymmetry $A^P_C$ was consistent with 
zero.  \babar has recently performed a measurement of all twelve asymmetries [12,13].
LHCb with its large sample of data has provided an interesting insight into these
decays, as they have been studied in bins of $K^+K^-$ and $\pi^+\pi^-$ invariant
masses as well as performing phase space integrated measurements [14].  
The distributions for these invariant mass distributions indicate a rich
resonant structure in the final state. 
In addition to the $K^+K^-$ and $\pi^+\pi^-$ combination studied one should investigate the 
$K^\pm\pi^\mp$ combinations to facilitate building a robust amplitude model to further study of the data.  
Interpretation of these results is complicated by the lack of a detailed amplitude model, however
by considering the results of the simple model discussed below one can conclude that
there is no evidence for a non-zero weak phase difference in this decay.  All of the non-zero
asymmetries measured by \babar and LHCb can be driven by strong phase differences (see Section 2.4).

The channel $D^+ \to K_S K^+ \pi^+\pi^-$ has been studied by \babar, where all measured asymmetries
are found to be consistent with zero (integrating over phase space) [15].  The branching fraction
for this channel is $(1.75 \pm 0.18)\times 10^{-3}$. It remains to be
seen if there is a more complex picture that averages out to this null result when integrating
over phase space.  The corresponding $D_s^+$ decay has a branching fraction of $(1.03 \pm 0.10)\times 10^{-3}$
and also been studied [15].  
Here the pattern observed for $D^0 \to K^+K^-\pi^+\pi^-$ is repeated; the asymmetries
driven by a non-zero weak phase difference are all zero, but those that can be driven by
strong phase differences are not.   The decays $D^+_{(s)} \to K_L K^+ \pi^+\pi^-$ have not
been studied by LHCb or the $B$ factories.  Given the presence of the $K_L$ in the final
state one would expect a small residual level of \CP violation from kaon decays to be present.
These modes would have significant amounts of background at a $B$ factory and would be difficult to
attempt to reconstruct in a hadronic environment like the LHC.  A significant virtue of a $\tau$-charm factory is
the ability to infer the missing energy and effectively reconstruct the $K_L$ four momentum.
 With 1\invab of data one could make precision measurements of triple product asymmetries in $D^+ \to K_{S,L} K^+ \pi^+\pi^-$. 
A data sample of 100\invfb collected at $D_s$ threshold would provide about $32\times 10^6$
$D^\pm_s$ to perform similar measurements.  This would be sufficient to provide several tens of thousands of
$D^+_s \to K_{S,L} K^+ \pi^+\pi^-$ decays to study.  A statistical precision on triple product 
asymmetries better than a percent would be achievable with such a sample.

%
%
A $\tau$-charm factory is well placed to perform precision measurements of these, and many other decay
channels.  For final states with one or more neutral meson there are obvious advantages in using
data from an \epem environment compared with $pp$ collisions at the LHC.  Kang and Li have studied
the prospects for a variety of $D$ decays to $VV$ final states [16] (here $V$ is a vector particle with $J^P=1^-$).  Sub-percent level precisions
are attainable with modest data samples ($\sim 20\invfb$) by BES II for the modes studied: neutral
$D$ meson decays to $\rho^0\rho^0$, $\overline K^{*0}\rho^0$, $\rho^0\phi$, $\rho^+\rho^-$, $K^{*+}K^{*-}$ and $K^{*0}\overline{K}^{*0}$,
and charged $D$ decays to $\overline{K}^{*0} \rho^+$.  The statistical precision of the 
charged $D$ decay is at the per mille level with this sample size. A super $\tau$-charm factory would be expected to 
accumulate significantly larger samples of data than this.  
For example a factory accumulating 
$1\invab$ of data at charm threshold could achieve statistical uncertainties on triple product 
asymmetry measurements for all of these decays at or below the per mille level.

%
%
If we consider the kaon measurements discussed in 2.1, these are triple product asymmetries from four body decays
derived from \CP violating and \CP conserving two body decays of the kaon.  The equivalent possibility for 
investigation in charm has been ignored thus far; i.e. the search for \CP violation in 
$D$ decays to $h^+h^- \ell^+\ell^-$ final states, where $h=K, \pi$ and $\ell = e, \mu$.  
Assuming that $D\to K^+K^+$ and/or $\pi^+\pi^+$ would exhibit \CP violation
at some level, one could use the interference between amplitudes generated in an analogous way to generate
an asymmetry in these decays.   The PDG reports upper limits on these modes ranging between $3.1\times 10^{-4}$
and $3.0\times 10^{-5}$ [17].  Some of these limits are just above naive expectations of
the branching fractions based on the known two body final state branching fractions.  The first step would be
to search for these data at a $B$, $\tau$-charm factory or the LHC and subsequently explore the triple product
asymmetry structure of the decays to search for symmetry violation.  It is worth noting that an advantage of 
these modes is that they are unambiguous; the hadronic and di-lepton systems can be treated as $ab$ and $cd$,
respectively; unlike the current measurements where there are two possible pairing combinations to consider
when probing amplitudes.  The corresponding set of measurements for $D_{(s)}^+$ decays would involve 
$h^+ h^{\prime 0}\ell^+\ell^-$ final states, where $h=\pi$, $K$.  The Cabibbo suppressed decays would allow us to search
for \CP violation, and the Cabibbo favoured states would provide useful control samples; however 
$K_{L,S}\pi^+ \ell^+\ell^-$ states would ultimately have a small \CP violating effect resulting
from the kaon \CP violation intrinsic to the final state.  
A number of four body \Dz and $D^\pm_{(s)}$ decays are yet to be studied; it is worth noting that the
modes measured so far all have large branching fractions.  Rare decays are more suitable for searches
for physics beyond the SM as small SM amplitudes can generate large effects when beating against
any hypothetical new physics amplitude of a comparable size.  The one thing that we do know about 
new physics amplitudes is that they are at best small for the energy scale being probed.
Thus multi-body rare charm decays may provide an interesting 
test bed for \CP violation; In addition to obtaining a more complete understanding of the copious decays,
experimentalists should study the available data for the rarer processes. 
It remains to be seen if one can generate large effects in the SM 
in analogy with the $K_L\to \pi^+\pi^-e^+e^-$ case.

%
%
\vspace*{7mm} {\color{ooorangec}\hrule}\vspace{2mm} \noindent
{\color{ooorangec}\large
\usefont{T1}{fradmcn}{m}{n}\xbt 2.2.2\quad Charm baryon decays }\vspace{3.5mm}

While measurements so far have focused on mesons, there is also a rich area of study in the decay of charm baryons.
These systems are accessible using data from Belle II, BES III, the LHC, and a super $\tau$ charm facility.
The prospects for $\Lambda_c$ decays to final states including baryons, pseudoscalars and vector particles
have been studied in Ref. [18].  This paper assumes one year of data taking corresponding to an integrated
luminosity of 5 \invfb at the $X(4630)$ peak with BES-III.  This data sample corresponds to $2.5\times 10^6$ $\Lambda_c^+\Lambda_c^-$
pairs.  The estimated precisions attainable for 
triple product asymmetries with such a data sample are typically at the level of a few percent.  
A high luminosity $\tau$-charm facility would provide the opportunity to reach the sub-percent level in all modes
studied with a data sample of about 80 \invfb. If one considers the mode $\Lambda_c \to \Lambda(p\pi^-)\rho^+(\pi^+\pi^0)$, one could
reach a per mille level statistical precision on the triple product asymmetries with data samples as small as 100 \invfb.
As \CP violation is expected to be small in the charm sector these decays provide an excellent set of laboratories
to search for physics beyond the SM.  With 100 \invfb of data one would have about $5\times 10^{7}$ $\Lambda_c^+\Lambda_c^-$
pairs which would enable searches for rare decays of the $\Lambda_c^+$.

\vspace*{7mm} {\color{ooorangec}\hrule}\vspace{2mm} \noindent
{\color{ooorangec}\large
\usefont{T1}{fradmcn}{m}{n}\xbt 2.3\quad $\tau$ decays }\vspace{3.5mm}

Searches for \CP violation in $\tau$ decay have concentrated on the channel $\tau \to K_S\pi \nu$ [19,20].  The level
of experimental sensitivity is approaching that of the intrinsic effect of \CP violation in neutral kaons, which is
a SM background to the search for new physics.  One of the problems with performing a triple product asymmetry measurement
for a tau decay such as $\tau\to h h^\prime h^{\prime\prime}\nu$, where
$h^{(\prime[\prime])} = K$, $\pi$, $\eta$, is that the center of mass frame needs to be determined.  Here a $\tau$-charm 
factory has an advantage over other experimental facilities; while running on $\tau^+\tau^-$ threshold the leptons are
created at rest in the laboratory frame, and hence the kinematics are fully constrained by the observed four
momenta of the reconstructed particles.  Energy-momentum conservation allows one to infer the neutrino and hence
fully reconstruct the event.  In doing so it becomes possible to compute the full set of triple product asymmetries outlined at the 
start of this section in the search for new physics.  Decays with odd numbers of charged kaons in the final state
suffer from detection asymmetry effects which are well known, but provide additional systematic uncertainties.  Those
with neutral kaons suffer from regeneration and interference effects, which again provide additional 
uncertainties which come into play when interpreting results.  Higher energy systems may be able to 
perform triple product asymmetry measurements, however those are affected by the 
fact that it is not possible to fully reconstruct the decay for energies above threshold. 
The decays $\tau \to \pi^-\pi^0 K^0\nu$, $K^-\pi^0 K^0\nu$, and $\pi^-K^0\eta\nu$ are all expected to manifest 
\CP violation, resulting from the neutral kaon in the final state, and provide an 
interesting complement to the $\tau \to K_S\pi \nu$ mode already studied.  Any large \CP violation effect observed
in $\tau$ decay would be a clear sign of new physics.  This is a largely unexplored experimental area that
can be studied extensively at a $\tau$-charm facility such as BES III, or at a super $\tau$-charm factory.

\vspace*{7mm} {\color{ooorangec}\hrule}\vspace{2mm} \noindent
{\color{ooorangec}\large
\usefont{T1}{fradmcn}{m}{n}\xbt 2.4\quad A simple model }\vspace{3.5mm}

We can increase our understanding of the twelve triple product asymmetries introduced in Ref. [8] by considering
a simple model of two interfering scalar amplitudes divided into $+$ and $-$ parts according to the 
sign of the scalar triple product:
\begin{eqnarray}
A_+ &=& a_1 e^{i (\phi_1 + \delta_{1, +})} +  a_2 e^{i (\phi_2 + \delta_{2, +})},\\
A_- &=& a_1 e^{i (\phi_1 + \delta_{1, -})} +  a_2 e^{i (\phi_2 + \delta_{2, -})},\\
\overline A_+ &=& a_1 e^{i (-\phi_1 + \delta_{1, +})} +  a_2 e^{i (-\phi_2 + \delta_{2, +})},\\
\overline A_- &=& a_1 e^{i (-\phi_1 + \delta_{1, -})} +  a_2 e^{i (-\phi_2 + \delta_{2, -})},
\end{eqnarray}
where $\delta$ represents a strong phase and $\phi$ a weak phase.
Here the coefficients $a_1$ and $a_2$ are just the magnitudes of the interfering amplitudes.
In this case, as shown in Ref. [8], the six asymmetries $A^P_C$, $A_C$, $\overline A_C$, $A^C_P$, $A^\C_{\CP}$, and $A^{\CP}_{\C}$
can only be non-zero if the difference between the weak phases is non-zero.  The remaining asymmetries can be non-zero even if
$\phi_1 - \phi_2 = 0$.  This simple model can be extended from the interfering (pseudo)scalar amplitude case to a more general
scenario amplitudes with higher spins following the procedure outlined in [21].

%
%
\vspace*{7mm} {\color{ooorangec}\hrule}\vspace{2mm} \noindent
{\color{ooorangec}\large
\usefont{T1}{fradmcn}{m}{n}\xbt 3\quad Tests using entangled states}\vspace{3.5mm}

\noindent 
John Bell resolved the EPR conundrum in 1961, and in doing so invented the concept of entangled quantum states [22].  
$\epem$ collisions at a centre of mass energy of 3770 \mev, corresponding to the $\psi(3770)$ allow us to 
prepare quantum correlated pairs of neutral $D$ mesons.  
In analogy with the Stern-Gerlach experiment, any pair of orthonormal states can be used to describe the system.
It is convenient to use quark flavour $\{D^0, \overline{D}^0 \} \equiv \{\ell^+X, \ell^-X \}$ and 
\CP eigenstates $\{D_+, D_- \} \equiv \{+1, -1\}$ to write the wave function:
\begin{eqnarray}
\Psi &=& \frac{1}{\sqrt{2}} \left( D_1^0\overline{D}_2^0 - \overline{D}_1^0 D_2^0 \right),\\
     &=& \frac{1}{\sqrt{2}} \left( D_{1, +} D_{2, -} - D_{1, -} D_{2, +} \right).
\end{eqnarray}
The subscripts $\pm$ denote the \CP eigenvalue of the $D$ decay as even or odd, respectively.  The Roman numeral
subscripts refer to the time ordering of decaying mesons; either the first (1) or second (2) meson to decay.  The 
second set written down indicates the final state reconstructed in for the flavour basis, or \CP eigen value for the 
\CP basis.  The filter decays to a lepton $+ X$ are an accurate way to determine the quark flavour in a charm decay,
the mis-tag probability at an \epem machine running at charm threshold is small.  The set of \CP filter
decays to complement these include $\eta_{\CP} = +1$ ($-1$)  $D\to h^+h^-$ where $h=\pi$, $K$ 
and $D\to K_L \omega (\to \pi^+\pi^-\pi^0)$ and $K_L\phi(\to K^+K^-)$
($D\to K_S \omega (\to \pi^+\pi^-\pi^0)$ and $K_S\phi(\to K^+K^-)$).

We can consider the possible combinations of decay to occur via either the flavour or \CP filters described above, 
which gives rise to three possible measurements of interest.  However, it is 
useful to note that in addition to filtering using only flavour or only \CP states, we can also filter using
a combination flavour then \CP filters or \CP then flavour filters.  This results in a total of 15 distinct
asymmetries [23] as listed in Table 1.  
The two flavour filter only asymmetries have been studied for many decades.  The \CP filter only asymmetry has not
been studied before for any neutral meson system.  
The remaining twelve asymmetries are derived using the approach 
described in Refs [24,25], and measured by \babar\ for
neutral $B$ decays [26].  It is worth noting that when using only a single filter basis pair it is 
not possible to construct an unambiguous test of a single symmetry, however the constructed asymmetry can be
used to simultaneously test a pair of symmetries.  When using two filter basis pairs it is possible to 
resolve the remaining ambiguity to obtain a set of tests of only one symmetry.

In general one should perform these measurements as a function of the proper time difference between the first
and second $D$ meson decays in the event (usually denoted as $\mathrm \Delta t$ in the literature, for example see
Refs [4,5] for details of time-dependent analyses).  However the mixing frequency and lifetime difference 
between \Dz and \Dzb is small in the charm system; $x = \mathrm \Delta m / \Gamma \sim 0.5\%$ and $y=\mathrm \Delta \Gamma / 2 \Gamma \sim 0.7\%$.
Hence initially time-integrated measurements of the asymmetries outlined below would be of direct 
interest; and a small correction would be required when interpreting precision measurements 
in order to take into account the fact that $x$ and $y$ are non-zero.

{\small
Table 1: The fifteen possible pairings of reference and symmetry conjugated transitions used to study \CP, \T and \CPT for
pairs of neutral $D$ mesons.\\
\begin{center}
\begin{tabular}{ccc} \hline\hline
Symmetry         & Reference & Conjugate \\ \hline
\CP and \T       & $\Dz\to \Dzb$ & $\Dzb\to \Dz$  \\
\CP and \CPT     & $\Dz\to \Dz$  & $\Dzb\to \Dzb$ \\ \hline
\T and \CPT      & $D_+\to D_-$ & $D_-\to D_+$  \\ \hline
\CP              & $\Dzb\to D_-$ & $\Dz \to D_-$  \\
                 & $D_+\to \Dz$  & $D_+ \to \Dzb$ \\
                 & $\Dzb\to D_+$ & $\Dz\to D_+$   \\
                 & $D_-\to \Dz$  & $D_-\to \Dzb$  \\ \hline
\T               & $\Dzb\to D_-$ & $D_- \to \Dzb$ \\
                 & $D_+\to \Dz$  & $\Dz \to D_+$  \\
                 & $\Dzb\to D_+$ & $D_+\to \Dzb$  \\
                 & $D_-\to \Dz$  & $\Dz\to D_-$   \\ \hline
\CPT             & $\Dzb\to D_-$ & $D_- \to \Dz$  \\
                 & $D_+\to \Dz$  & $\Dzb \to D_+$ \\
                 & $\Dz\to D_-$  & $D_-\to \Dzb$  \\
                 & $D_+\to \Dzb$ & $\Dz\to D_+$   \\ \hline\hline
\end{tabular} \label{tbl:appendix_asymmetries:combinations:fifteeen}
\end{center}
}

Section 3.1 discusses measurements of asymmetries constructed from the flavour filter basis pair, 
Section 3.2 discusses possible measurements of the asymmetry constructed from \CP filter basis
pairs, and Section 3.3 discusses the remaining measurements using a combination of 
\CP and flavour filter basis pairs.

\vspace*{7mm} {\color{ooorangec}\hrule}\vspace{2mm} \noindent
{\color{ooorangec}\large
\usefont{T1}{fradmcn}{m}{n}\xbt 3.1\quad Using flavour filters}\vspace{3.5mm}

\noindent
It is possible to construct tests of \CP and \T and of \CP and \CPT using 
flavour filter states.  These measurements require studies as
a function of lifetime difference between opposite and same sign tagged
final states.  The asymmetries that one measures are
\begin{eqnarray}
A_{\CP, \T}   &=& \frac{\Gamma(\Dz\to \Dzb) - \Gamma(\Dzb\to \Dz) }  {\Gamma(\Dz\to \Dzb) + \Gamma(\Dzb\to \Dz)},\\
A_{\CP, \CPT} &=& \frac{\Gamma(\Dz\to \Dz) - \Gamma(\Dzb\to \Dzb) }  {\Gamma(\Dz\to \Dz) + \Gamma(\Dzb\to \Dzb)}.
\end{eqnarray}
The former measurement is usually referred to as a measurement of \CP in mixing, however it is worth noting 
that this is also simultaneously testing \T, c.f. the Kabir asymmetry measured by CPLEAR in kaon decays [27].  
The typical experimental signature that one would pursue for this would be to reconstruct both $D$ mesons via a semi-leptonic
decay and search for same sign di-leptons; one being from each decay. A non-zero value of the 
resulting asymmetry $A_{\CP, \T}$ as a function of proper time difference between the decaying $D$ mesons 
would indicate a violation of both \CP and \T. The corresponding test for $A_{\CP, \CPT}$ requires opposite
sign dilepton final state, and a non-zero value of this asymmetry would indicate a violation of both \CP and \CPT.
This could only be manifest by physics beyond the SM.

It is worth noting that while these tests are performed using an entangled state prepared in the decay of
a $\psi(3770)$, it is also possible to use a hadronic production environment with associated production of charm 
to flavour tag the neutral $D$ meson at the point of production, and reconstruct the semileptonic decay at
a later time.  A second route that is viable at the LHCb experiment is to use semileptonic $B$ decays to
tag the flavour of the decaying neutral $D$ meson at the point of production, and the leptonic charge 
at the point of decay to provide the required rates to compute $A_{\CP, \T}$ and $A_{\CP, \CPT}$.

Over the past few years there has been a lot of interest in the like-sign semileptonic asymmetry measurement 
made by the D0 experiment for $B_s$ mesons [28].  This is a measurement of $A_{\CP, \T}$ using $B_s$ decays.
The reported D0 result is $A_{\CP, \T} = -0.787 \pm  0.172 \pm 0.093$, which deviates from the SM expectation
of zero by $3.9\sigma$.  
All corresponding measurements made by the $B$ factories for this asymmetry in $B_d$ mesons
are consistent with zero (See [4] and references therein).  
If the anomalous like-sign di-muon asymmetry in D0 is the result of some kind of new physics then that may also 
be manifest in the charm sector.
Hence, it is important to study charm decays in order to search for
evidence of \CP and \T violation.
As noted in [29] it is possible for systems with $\Delta \Gamma \simeq 0$ to result in a
zero asymmetry measurement for $A_{\CP, \T}$ even when the symmetry is violated.  For neutral charm (like $B_s$) mesons
$\Delta \Gamma \neq 0$; hence such a measurement for \Dz mesons is an important test to complement the studies performed thus far.

A recent review of semi-leptonic (SL) decays by Lui outlines experimental issues related to reconstructing these states [30].  The 
branching fraction of SL decays is large, and so precision measurements of $A_{\CP, \T}$ and $A_{\CP, \CPT}$ are in principle
achievable assuming that systematic uncertainties may be kept under control.
\clearpage

\vspace*{7mm} {\color{ooorangec}\hrule}\vspace{2mm} \noindent
{\color{ooorangec}\large
\usefont{T1}{fradmcn}{m}{n}\xbt 3.2\quad Using \CP filters}\vspace{3.5mm}

\noindent
The asymmetry
\begin{eqnarray}
A_{\T, \CPT}   = \frac{\Gamma(D_+\to D_-) - \Gamma(D_-\to D_+) }  {\Gamma(D_+\to D_-) + \Gamma(D_-\to D_+) },
\end{eqnarray}
constructed using only \CP filter states allows us to perform a simultaneous test of both \T and \CPT.  In order
to perform this test we need to identify $D$ meson decays into \CP even and \CP odd final states.
For example one can measure
the asymmetry between $D\to K_S (\omega, \phi, \rho^0)$ followed by $D\to h^+h^-$ (or $D\to K_L (\omega, \phi, \rho^0)$) 
and $D\to h^+h^-$ (or $D\to K_L (\omega, \phi, \rho^0)$) followed by 
$D\to K_S (\omega, \phi, \rho^0)$ final states. Any combination of $+1$ and $-1$ states can be used to test \T
using this method.  The SM expectation is that $A_{\T, \CPT} = 0$.  Any non-zero value 
for any of these combinations would indicate violation of both \T and \CPT, and physics beyond the SM.  
This type of test complements the flavour filter tests of $A_{\CP, \T}$ and $A_{\CP, \CPT}$ described above.
The initial \CP filter state can be tagged via the decay of a $\psi(3770)$.   As a result of incoherent 
production of charm at a hadron collider or $B$ factory does not permit an obvious route to performing this type 
of asymmetry measurement via other means.

Experimentally the double $D\to K_{S,L} \omega^0$ decays should proceed with a rate of the order of $1.2\times 10^{-4}$.
Allowing for the ability to reconstruct these decays with a modest efficiency would set the single event sensitivity 
at a the level of $O(few\, 10^{-5})$.  A Super $\tau$-charm factory would be able to accumulate about 10000 events with
1\invab to perform a measurement of this type. The double decays to $D\to K_{S,L} \rho^0$ and $D\to K_{S,L} \phi$ have
product branching fractions of $3.6\times 10^{-5}$ and $4\times 10^{-6}$, respectively.  Samples of about
1000 and 100 events, respectively could be recorded in order to permit a measurement of $A_{\T, \CPT}$ for these decays.

\vspace*{7mm} {\color{ooorangec}\hrule}\vspace{2mm} \noindent
{\color{ooorangec}\large
\usefont{T1}{fradmcn}{m}{n}\xbt 3.3\quad Using both flavour and \CP filters}\vspace{3.5mm}

\noindent
The remaining twelve asymmetries can be constructed from Table 1 and these constitute 
four tests of each of \CP, \T and \CPT.  These tests complement the $A_{\CP, \T}$, $A_{\CP, \CPT}$ and
$A_{\T, \CPT}$ asymmetries discussed above as they each unambiguously identify one symmetry to test.
These asymmetries have only been measured thus far for neutral $B$ mesons [26], where results
consistent with the SM were obtained; namely that \CP and \T are violated, whilst \CPT remains conserved.
These measurements provide an important cross check of our understanding of symmetry violation
to complement existing routes to search for symmetry violation.  The magnitudes of asymmetries 
determined in these decays are related to unitarity triangle angles in the charm sector
(just as the asymmetries measured in Ref. [26] are related to $\sin 2\beta$ from the $B_d$ ``Unitarity Triangle'').  
As \CP violation is expected to be small in the charm sector, so the angles measurable in the \CP and \T 
asymmetries are expected to be small (i.e. compatible with zero within uncertainties).  
The \CPT asymmetries are expected to be zero in the SM, to signify that this symmetry is conserved.
Significant deviations from this pattern would be an indication of physics beyond the SM.
A discussion of how to relate the angles of the charm unitarity triangle to decays in the 
charm system can be found in Ref. [5].

Table 2 summarises the fifteen asymmetries in terms of the final states that must be reconstructed
for reference and conjugated processes.  These clearly highlight the symmetries being tested by
``same'' and ``opposite sign'' asymmetry measurements, as well as allowing one to clearly identify
the combinations for testing the remaining thirteen quantities.

{\small
Table 2: Final states reconstructed for the first and second $D$ mesons in an event, along with the
conjugate processes to test the symmetries \CP, \T and \CPT. The $\ell^+X$ ($\ell^-X$) state is the flavour 
filter for a $\Dz$ (\Dzb), and $+1$ and $-1$ indicate the \CP filter decays listed in the text.\\
\begin{center}
\begin{tabular}{ccc} \hline\hline
Symmetry         & Reference & Conjugate \\ \hline
\CP and \T       & $(\ell^-X, \ell^-X)$  & $(\ell^+X, \ell^+X)$  \\
\CP and \CPT     & $(\ell^-X, \ell^+X)$  & $(\ell^+X, \ell^-X)$ \\ \hline
\T and \CPT      & $(-1, -1)$ & $(+1, +1)$  \\ \hline
\CP              & $(\ell^-X, -1)$ & $(\ell^+X , -1)$  \\
                 & $(+1, \ell^+X)$  & $(+1 , \ell^-X)$ \\
                 & $(\ell^-X, +1)$ & $(\ell^+X, +1)$   \\
                 & $(-1, \ell^+X)$  & $(-1, \ell^-X)$  \\ \hline
\T               & $(\ell^-X, -1)$ & $(-1 , \ell^-X)$ \\
                 & $(+1, \ell^+X)$  & $(\ell^+X , +1)$  \\
                 & $(\ell^-X, +1)$ & $(+1, \ell^-X)$  \\
                 & $(-1, \ell^+X)$  & $(\ell^+X, -1)$   \\ \hline
\CPT             & $(\ell^-X, -1)$ & $(-1 , \ell^+X)$  \\
                 & $(+1, \ell^+X)$  & $(\ell^-X , +1)$ \\
                 & $(\ell^+X, -1)$  & $(-1, \ell^-X)$  \\
                 & $(+1, \ell^-X)$ & $(\ell^+X, +1)$   \\ \hline\hline
\end{tabular} \label{tbl:appendix_asymmetries:combinations:final}
\end{center}
}

%
%
\vspace*{7mm} {\color{ooorangec}\hrule}\vspace{2mm} \noindent
{\color{ooorangec}\large
\usefont{T1}{fradmcn}{m}{n}\xbt 3.4\quad Tests of quantum mechanics}\vspace{3.5mm}

A natural question to ask when presented with an entangled system 
is ``can one test Bell's inequality using entangled neutral mesons?''.  The Belle experiment
has attempted to address this issue in the context of $B_d^0\overline{B}_d^0$ mesons [4], however there are limitations
of that approach that prohibit this possibility. Those limitations also preclude the possibility of a test of 
quantum mechanics using neutral $D$ mesons [31] given that $x$ is small.   However, 
it may be possible to test for decoherence effects in the entangled wave function in analogy with measurements performed by 
the $B$ factories.  The neutral charm system provides several experimental advantages over $B$ mesons for this kind of
test; for example flavour tagging can be performed with essentially no dilution on the precision (and hence minimal systematic
uncertainty) of the flavour assignment.  The small magnitude of mixing for charm may also prove to be advantageous for 
such a test. For a discussion of decoherence tests see for example Ref. [32].

%
%
\vspace*{7mm} {\color{ooorangec}\hrule}\vspace{2mm} \noindent
{\color{ooorangec}\large
\usefont{T1}{fradmcn}{m}{n}\xbt 4\quad Summary  }\vspace{3.5mm}

\noindent 
A high luminosity $\tau$-charm factory 
would allow a number of interesting measurements to be performed.  It will be possible to 
explore discrete symmetry (non-)conservation in charm meson decays using entangled neutral
$D$ mesons created via decays of the $\psi(3770)$ resonance, and to explore \C, \P, and 
\CP violation in $\tau$ lepton, charm meson and baryon decay.  There are advantages of
performing such measurements at an \epem collider over other facilities, in particular tests
of the full set of possible \T and \CPT asymmetries require the use of entangled pairs of 
neutral $D$ mesons, unique to a $\tau$-charm factory.  
A number of triple product asymmetry measurements are discussed in the context of searching 
for discrete symmetry violation.  Half of these measurements are tests of a
non-zero weak phase difference (related to the phase of the CKM matrix).  One can also use
entangled states to study \CP, \T and \CPT symmetries. 
At this time it is not clear what the
best way to discover \CP violation in the charm sector is; as a result one should perform
all possible measurements that may lead to an effect.  At the same time it is important to 
perform tests of the other discrete symmetries in the hope of further elucidating our understanding
of the SM of particle physics.  While it is not possible to perform a test of Bell's inequalities with
charm mesons, it will be possible to search for decoherence of the wave function for 
entangled pairs of neutral $D$ mesons.

\vspace{20mm}

{\color{ooorangec}\hrule}\vspace{2mm} \noindent
{\color{ooorangec}\large
\usefont{T1}{fradmcn}{m}{n}\bf  References and notes }\vspace{-2mm}

\parskip=0mm \baselineskip 15pt\renewcommand{\baselinestretch}{1.25} \footnotesize

\begin{OOExercises}
%
%
\item C.~S.~Wu, E.~Ambler, R.~W.~Hayward, D.~D.~Hoppes and R.~P.~Hudson,
  Phys.\ Rev.\  {\bf 105}, 1413 (1957).

\item J.~H.~Christenson, J.~W.~Cronin, V.~L.~Fitch and R.~Turlay,
  Phys.\ Rev.\ Lett.\  {\bf 13}, 138 (1964).

\item For example see P. K. Kabir, 
 The\CP Puzzle, Academic Press (1968) and R. G. Sachs, The physics of time reversal, University of Chicago Press (1987).

\item A.~J.~Bevan {\it et al.} [BaBar and Belle Collaborations],
  Eur.\ Phys.\ J.\ C {\bf 74} (2014) 3026
  [arXiv:1406.6311 [hep-ex]].

\item A.~J.~Bevan, G.~Inguglia and B.~Meadows,
  Phys.\ Rev.\ D {\bf 87} (2013) 3,  039905
   [Phys.\ Rev.\ D {\bf 87} (2013) 3,  039905]
  [arXiv:1106.5075 [hep-ph]].

\item M.~Gersabeck,
  PoS FWNP {\bf } (2015) 001
  [arXiv:1503.00032 [hep-ex]].

%
%

%
%
\item
  M.~Gronau and J.~L.~Rosner,
  Phys.\ Rev.\ D {\bf 84} (2011) 096013
  [arXiv:1107.1232 [hep-ph]].

\item
  A.~J.~Bevan,
  arXiv:1408.3813 [hep-ph].

%
%
\item 
   P.~Heiliger and L.~M.~Sehgal,
   Phys.\ Rev.\ D {\bf 48} (1993) 4146
   [Phys.\ Rev.\ D {\bf 60} (1999) 079902].

\item
  A.~Alavi-Harati {\it et al.} [KTeV Collaboration],
  Phys.\ Rev.\ Lett.\  {\bf 84} (2000) 408
  [hep-ex/9908020];   
  E.~Abouzaid {\it et al.} [KTeV Collaboration],
  Phys.\ Rev.\ Lett.\  {\bf 96} (2006) 101801
  [hep-ex/0508010].

\item 
  A.~Lai {\it et al.} [NA48 Collaboration],
  Eur.\ Phys.\ J.\ C {\bf 30} (2003) 33.

\item
  J.~P.~Lees {\it et al.} [BaBar Collaboration],
  Phys.\ Rev.\ D {\bf 84} (2011) 031103
  [arXiv:1105.4410 [hep-ex]].

\item
  M. Martinelli, contribution to the 8th international work-
  shop on the CKM Unitarity Triangle, Vienna (2014).

\item 
  R.~Aaij {\it et al.} [LHCb Collaboration],
  JHEP {\bf 1410} (2014) 005
  [arXiv:1408.1299 [hep-ex]].

\item
  J.~P.~Lees {\it et al.} [BaBar Collaboration],
  Phys.\ Rev.\ D {\bf 84} (2011) 031103
  [arXiv:1105.4410 [hep-ex]].

\item 
  X.~W.~Kang and H.~B.~Li,
  Phys.\ Lett.\ B {\bf 684} (2010) 137
  [arXiv:0912.3068 [hep-ph]].

\item
  K.A. Olive et al. (Particle Data Group), Chin. Phys. C, {\bf 38}, 090001 (2014).

\item 
  X.~W.~Kang, H.~B.~Li, G.~R.~Lu and A.~Datta,
  Int.\ J.\ Mod.\ Phys.\ A {\bf 26} (2011) 2523
  [arXiv:1003.5494 [hep-ph]].

%
%
\item 
  M.~Bischofberger {\it et al.} [Belle Collaboration],
  Phys.\ Rev.\ Lett.\  {\bf 107} (2011) 131801
  [arXiv:1101.0349 [hep-ex]].

\item
  J.~P.~Lees {\it et al.} [BaBar Collaboration],
  Phys.\ Rev.\ D {\bf 85} (2012) 031102
   [Phys.\ Rev.\ D {\bf 85} (2012) 099904]
  [arXiv:1109.1527 [hep-ex]].

%
%
\item
  G.~Valencia,
  Phys.\ Rev.\ D {\bf 39} (1989) 3339.

%
%

\item 
  J. S. Bell, Physics {\bf 1}, 195 (1964).

\item 
  A.~Bevan,
  arXiv:1505.06943 [hep-ex].

\item
  M.~C.~Banuls and J.~Bernabeu,
  Nucl.\ Phys.\ B {\bf 590} (2000) 19
  [hep-ph/0005323].

\item
  J.~Bernabeu, F.~Martinez-Vidal and P.~Villanueva-Perez,
  JHEP {\bf 1208} (2012) 064
  [arXiv:1203.0171 [hep-ph]].

\item
  J.~P.~Lees {\it et al.} [BaBar Collaboration],
  Phys.\ Rev.\ Lett.\  {\bf 109} (2012) 211801
  [arXiv:1207.5832 [hep-ex]].

\item 
  A. Angelopoulos et al. [CPLEAR Collaboration], Phys. Lett. B {\bf 444}, 43-51 (1998).

\item 
  V.~M.~Abazov {\it et al.} [D0 Collaboration],
  Phys.\ Rev.\ D {\bf 82} (2010) 032001
  [arXiv:1005.2757 [hep-ex]].

\item 
 E. Alvarez and A. Szynkman, Mod. Phys. Lett. A {\bf 23}, 2085 (2008) [hep-ph/0611370].

%
%

\item
C.~Liu,
  arXiv:1302.0227 [hep-ex].

%
%

%
%

%
%

%
%

\item
R.~A.~Bertlmann, A.~Bramon, G.~Garbarino and B.~C.~Hiesmayr,
  Phys.\ Lett.\ A {\bf 332} (2004) 355
  [quant-ph/0409051].

\item
  R.~A.~Bertlmann, W.~Grimus and B.~C.~Hiesmayr,
  Phys.\ Rev.\ D {\bf 60} (1999) 114032
  [hep-ph/9902427].

\end{OOExercises}
\end{multicols}
\end{document}